\documentclass[preprint,preprintnumbers,amsmath,amssymb]{revtex4}
\usepackage{graphics,epsfig,subfigure,amsmath}
\usepackage{color}
\usepackage{braket}

\begin{document}
\renewcommand{\baselinestretch}{1.3}
\newcommand\beq{\begin{equation}}
\newcommand\eeq{\end{equation}}
\newcommand\beqn{\begin{eqnarray}}
\newcommand\eeqn{\end{eqnarray}}
\newcommand\nn{\nonumber}
\newcommand\fc{\frac}
\newcommand\lt{\left}
\newcommand\rt{\right}
\newcommand\pt{\partial}

\title{The Wigner function for Integer quantum Hall effect}
\author{Yao Wang$^1$\footnote{15591661215@163.com},
        Qi-Ming Fu$^1$\footnote{fuqiming@snut.edu.cn},
        Huawei Fan$^1$,
        Li Yu$^1$,
        Kang Li$^2$}.
 \affiliation{$^1$ Shaanxi University of Technology, Han zhong 723001, China \\
              $^2$ School of Physics, Hangzhou Normal University, Hangzhou, Zhejiang 311121, China}

\begin{abstract}
  Wigner's quasi-probability distribution function in phase space is a specialized representation of the density matrix, possessing significant physical importance. In this article, we first review the wave function describing electronic motion in an electromagnetic field under the Landau gauge. Next, based on an introduction to the properties of the Wigner function, we calculate the Wigner function for the integer quantum Hall effect using the integral method.
\end{abstract}

\maketitle

\section{Introduction}

In 1932, the Wigner function was introduced into quantum mechanics as a quasi-probability distribution function in phase space. It represents a specialized formulation of the density matrix and holds significant physical importance \cite{1}. The Wigner function not only plays a critical role in quantum optics, nuclear physics, and quantum chaos but also serves as an effective semiclassical approximation. Its applications extend widely to quantum computation, quantum information, and signal processing. However, the advantages of the Wigner function were not fully recognized until the 1970s. In 1975, Moyal unveiled its potential as a compelling quantization method \cite{2}, derived from the internal logic of quantum mechanics. This method is equivalent to established quantization approaches¡ªsuch as Schrodinger, regularization, Heisenberg operator, and Feynman path integral quantization¡ªand is based on the Moyal star eigenvalue equation. Unlike traditional approaches, this logically complete and independent quantization framework does not require the selection of a specific representation space, such as coordinate or momentum representation. Furthermore, the Wigner function in phase space inherently incorporates the uncertainty relation. More importantly, its significance has gained prominence in modern quantum measurement. Specifically, the Wigner function, as a real-valued function in phase space, exhibits the properties of a quasi-probability distribution function. For instance, in Ref.~\cite{3}, the Wigner function for an atomic beam in a double-slit interference experiment was measured, with results consistent with theoretical predictions.

In 1879, Klitzing, Dorda, and Pepper discovered the integer quantum Hall effect, a groundbreaking finding that earned Klitzing the Nobel Prize in Physics in 1985 \cite{4,5,6,7,8,9,10}. To observe the quantum Hall effect, it is essential to confine electrons to two-dimensional motion, forming what is known as a two-dimensional electron gas. Under these conditions, the energy states of the electron gas exhibit discrete separations, corresponding to the Landau energy levels. These levels, first derived by Landau in 1930 while solving the problem of charged particles moving in a uniform magnetic field, are characterized by their highly notable feature of infinite degeneracy. The integer quantum Hall effect is typically understood within the framework of a single-electron model, which provides critical insights into this phenomenon.

The paper is organized as follows: the first section is the introduction, which shows the importance of integer quantum hall effect and landau energy levels. In Sec.~\ref{sec2}, we review the wave function for moving electrons in the electromagnetic field. In Sec.~\ref{sec3}, we introduce the important properties of the Wigner function. In Sec.~\ref{sec4}, we obtained the corresponding Wigner function by the method of integral. Section \ref{sec6} comes with the conclusion.

\section{A review of electronic motion wave function in the electromagnetic field }~\label{sec2}

The quantum Hall effect is a quantum mechanical counterpart of the classical Hall effect, where the principles of quantum mechanics are essential for its study. As a result, the observed phenomena differ significantly from those predicted by classical Hall effect theory. The quantum Hall effect exhibits many intriguing phenomena that cannot be explained by classical theory.

The quantum Hall effect is categorized into the integer quantum Hall effect (IQHE) and the fractional quantum Hall effect (FQHE). The IQHE can be understood within the framework of a single-electron model, which is the focus of this article and will be described in detail. In contrast, the FQHE arises from strong electron-electron interactions, making it a prominent topic in condensed matter physics research. Beyond these, other variants such as the spin quantum Hall effect and the anomalous quantum Hall effect also represent active and significant areas of investigation in modern physics.

Now let us review the integer quantum hall effect. First we choose Landau specification:
$A_x  =  - By\quad  A_y  = 0\quad A_z  = 0$, and $\varphi  =  - \varepsilon y$ of y direction, then the schrodinger equation is
\begin{equation}
i\hbar \frac{\partial }{{\partial t}}\psi  = \left[ {\frac{1}{{2\mu }}\left( { - i\hbar \frac{\partial }{{\partial x}} - \frac{{eB}}{c}y} \right)^2  + \frac{1}{{2\mu }}\left( { - i\hbar \frac{\partial }{{\partial y}}} \right)^2  + e\varepsilon y} \right]\psi,~\label{eq1}
\end{equation}
where $\mu$ and $e$ stand for electron mass and electronic charge, respectively. Eigenstate of $H$ desirable for conservation quantity complete set ($H, P_x$) simultaneous eigenstate, so the wave function can be written as
\begin{equation}
\psi \left( {x,y,t} \right) = e^{{{ip_x x} \mathord{\left/
 {\vphantom {{ip_x x} \hbar }} \right.
 \kern-\nulldelimiterspace} \hbar }} e^{{{iEt} \mathord{\left/
 {\vphantom {{iEt} \hbar }} \right.
 \kern-\nulldelimiterspace} \hbar }} \phi \left( y \right),~\label{eq2}
\end{equation}
where $p_x$ is the eigenvalue of $\hat p_x$. Inserting Eq.~(\ref{eq2}) into Eq.~(\ref{eq1}), one can obtain
\begin{equation}
\frac{1}{{2\mu }}\left[ {\left( {p_x  - \frac{{eB}}{c}y} \right)^2  - \hbar ^2 \frac{{d^2 }}{{dy^2 }} + e\varepsilon y} \right]\phi \left( y \right) = E\phi \left( y \right). ~\label{eq3}
\end{equation}

By defining $y_0 \equiv \frac{{cp_x }}{{eB}}$, $\omega  \equiv \frac{{eB}}{{\mu c}}$, Eq.~(\ref{eq3}) can be written as
\begin{equation}
 - \frac{{\hbar ^2 }}{{2\mu }}\phi ''\left( y \right) + \left[ {\frac{1}{2}\mu \omega ^2 \left( {y - y_0 } \right)^2  + e\varepsilon y} \right]\phi \left( y \right) = E\phi \left( y \right),
\end{equation}
which can be rearranged as
\begin{equation}
 - \frac{{\hbar ^2 }}{{2\mu }}\phi ''\left( y \right) + \frac{1}{2}\mu \omega ^2 \left( {y - y_0  + \frac{{e\varepsilon }}{{\mu \omega ^2 }}} \right)^2 \phi \left( y \right) = \left( {E + \frac{{e^2 \varepsilon ^2 }}{{2\mu \omega ^2 }} - e\varepsilon y_0 } \right)\phi \left( y \right).
\end{equation}
By defining $Y = y - y_0  + \frac{{e\varepsilon }}{{\mu \omega ^2 }}$ and $E' = E + \frac{{e^2 \varepsilon ^2 }}{{2\mu \omega ^2 }} - e\varepsilon y_0$, the above equation can be rewritten as
\begin{equation}
\left( { - \frac{{\hbar ^2 }}{{2\mu }}\frac{{d^2 }}{{dY^2 }} + \frac{1}{2}\mu \omega ^2 Y^2 } \right)\phi \left( Y \right) = E'\phi \left( Y \right),
\end{equation}
which is the same as one dimensional harmonic oscillator equation. The solution is
\begin{eqnarray}
\phi _n \left( y \right) &=& N_n \exp \left[ { - \frac{{\alpha ^2 }}{2}\left( {y - y_0  + \frac{{e\varepsilon }}{{\mu \omega ^2 }}} \right)} \right] H_n \left[ {\alpha \left( {y - y_0  + \frac{{e\varepsilon }}{{\mu \omega ^2 }}} \right)} \right], ~\label{eq7}\\
E'_n &=& \left( {n + \frac{1}{2}} \right)\hbar \omega , \quad n = 0,1,2,\cdots
\end{eqnarray}
with $\alpha=\sqrt{\mu\omega/\hbar}$. Then, the energy level $E_n$ can be obtained as
\begin{equation}
E_n  = E_n ^\prime   - \left( {\frac{{e^2 \varepsilon ^2 }}{{2\mu \omega ^2 }} - e\varepsilon y_0 } \right) = \left( {n + \frac{1}{2}} \right)\hbar \omega  - \frac{{e^2 \varepsilon ^2 }}{{2\mu \omega ^2 }} + \frac{{c\varepsilon p_x }}{B}.
\end{equation}

For the wave function $e^{{{iP_x x} \mathord{\left/
 {\vphantom {{iP_x x} \hbar }} \right.
 \kern-\nulldelimiterspace} \hbar }}$ of $x$  direction, since the $x$ direction is limited to the length of $L_x$, with the periodic boundary condition $\phi _{\rm{x}} ( - \frac{{L_x }}{2}) = \phi _{\rm{x}} (\frac{{L_x }}{2})$, we obtain

\begin{equation}
\phi _x (x) = \frac{1}{{\sqrt {L_x } }}\exp \left(\frac{{ip_x x}}{\hbar }\right),
\end{equation}
where $p_x  = \frac{{2\pi \hbar }}{{L_x }}n_x ,\quad n_x = 0,1,2,3 \cdots$.

Because of $y_0  = \frac{{cP_x }}{{eB}}$, and by observing the wave function, we all know $y_0$ is the center of the wave function, so $y_0$  can't be more than the boundary of the conductor. Ignoring the term of electric field, we can get $ - \frac{{L_y }}{2} \le y_0  \le \frac{{L_y }}{2}$. Plugging $y_0  = \frac{{cP_x }}{{eB}}$ into the equation, we obtain $- \frac{{eBL_y }}{{2c}} \le p_x  \le \frac{{eBL_y }}{{2c}}$. In addition, we also know that the momentum of $x$ direction is quantized,
i.e.,  $p_x  = \frac{{2\pi \hbar }}{{L_x }}n_x ,\quad n_x = 0,1,2,3 \cdots $. So we can get $ - \frac{{eBL_y L_x }}{{2hc}} \le n_x  \le \frac{{eBL_y L_x }}{{2hc}}$ . We can also take a maximum value of  $\frac{{eBL_y L_x }}{{2hc}}$ and minimum value of $ - \frac{{eBL_y L_x }}{{2hc}}$. Then the desirable number of possible
states in total is $\frac{{eBL_y L_x }}{{hc}}$ on each landau energy levels, and the number of per unit area state is $\frac{{eB}}{{hc}}$. From this we know when a landau level is just filled, the face of an electronic density is \cite{17}
\begin{equation}
n_s  = \frac{{eB}}{{hc}},
\end{equation}
which indicates $n_s$ is proportional to $B$.

\section{Expectations of the Wigner function and dynamics}~\label{sec3}

Quantization methods are typically classified into three main categories. The first method, developed in the 1920s by physicists such as Heisenberg, Schrodinger, and Dirac, is based on operator regularization within the framework of Hilbert space. The second method is the path integral formulation, introduced and formalized by Feynman. The third method is Moyal's multiplication regularization, which builds on the Wigner quasi-probability distribution function and the interplay between quantum mechanical operators and classical phase space \cite{11,12,13,14,15}.

As a quasi probability distribution function in the phase space, Wigner function is a good semi-classical approximation, which is very important in the physical measurement. It is known that in the phase space of freedom $s = n$, the general form of the Wigner function is \cite{16}:
\begin{equation}
W\left( {\vec x,\vec p;t} \right) = \frac{1}{{(2\pi )^n }}\int\limits_{ - \infty }^{ + \infty } {d\vec y^{} \exp \left( {i\vec y\vec p} \right)\left\langle {\vec x - \frac{\hbar }{2}\vec y} \right|\hat \rho \left| {\vec x + \frac{\hbar }{2}\vec y} \right\rangle }. ~\label{eq12}
\end{equation}

For the steady state, Eq.~(\ref{eq12}) becomes
\begin{equation}
W\left( {\vec x,\vec p} \right) = \frac{1}{{(2\pi )^n }}\int\limits_{ - \infty }^{ + \infty } {d\vec y^{} \varphi ^* (\vec x{\rm{ + }}\frac{\hbar }{2}\vec y)\exp \left( {i\vec y\vec p} \right)\varphi (\vec x - \frac{\hbar }{2}\vec y)}.
\end{equation}
The most important natures of Wigner function are
\begin{equation}
\int {d\vec x\;W(\vec x,\vec p)}  = \varphi ^ *  (\vec p)\,\varphi (\vec p) = \rho (\vec x),
\end{equation}
and
\begin{equation}
\int {d\vec p\;W(\vec x,\vec p)}  = \varphi ^ *  (\vec x)\,\varphi (\vec x) = \rho (\vec p).
\end{equation}

The Wigner function for space or the edge of the momentum space distribution is the probability density distribution of coordinate basis representation (i.e., momentum representation) . This is the very importance of the Wigner function.

In one dimensional case, the mechanical quantity  $f(x)$ is only relevant to coordinate. Thus its expectations can be expressed as

\begin{equation}
\left\langle {f(x)} \right\rangle  = \int\int {W(x,p)}f(x)dxdp.
\end{equation}
The mechanical quantity $g(p)$ is only related to the momentum, so its expectations can be written as
\begin{equation}
\left\langle {g(p)} \right\rangle  = \int\int {W(x,p)}g(p)dxdp.
\end{equation}
For the mechanics of $f(x) + g(p)$, its expectations can be expressed as follows
\begin{equation}
\left\langle {f(x) + g(p)} \right\rangle  = \int\int {W(x,p)}\left[ {f(x) + g(p)} \right]dxdp.~\label{eq18}
\end{equation}
Equation (\ref{eq18}) is the theoretical foundation for calculating expectations of current density with Hall electric field.

\section{Cross Wigner function} ~\label{sec4}

Next let us calculate the Wigner function for quantum hall effect. The Wigner in one dimensional space can be written as
\begin{equation}
W\left( {x,p_x } \right) = \frac{1}{{(2\pi )}}\int\limits_{ - \infty }^{ + \infty } {} dy\;\varphi ^* (x + \frac{\hbar }{2}y)\exp \left( {iyp_x } \right)\varphi (x - \frac{\hbar }{2}y). ~\label{eq31}
\end{equation}

Considering the above equation and formula (\ref{eq2}), we can work out the Wigner function of $x$ and $y$ direction respectively, so we can choose $W = W_x W_y$. In the $y$ direction, whereas $Y = y - y_0  + \frac{{e\varepsilon }}{{\mu \omega ^2 }}$, Eq.~(\ref{eq7}) will be changed into
\begin{equation}
\phi \left( Y \right) = N_n \exp \left[ { - \frac{{\alpha ^2 }}{2}\left( Y \right)^2 } \right]H_n \left[ {\alpha Y} \right]. ~\label{eq32}
\end{equation}

Then, inserting Eq.~(\ref{eq32}) into Eq.~(\ref{eq31}), we get
\begin{eqnarray}
W_n \left( {Y,p_y } \right) &=& \frac{{N_n^2 }}{{2\pi }}\int\limits_{ - \infty }^{ + \infty } {e^{ip_y x} H_n \left[ {\alpha \left( {Y + \frac{\hbar }{2}x} \right)} \right]H_n \left[ {\alpha \left( {Y - \frac{\hbar }{2}x} \right)} \right]} \nonumber\\
&&\exp \left\{ { - \frac{{\alpha ^2 }}{2}\left[ {\left( {Y + \frac{\hbar }{2}x} \right)^2  + \left( {Y - \frac{\hbar }{2}x} \right)^2 } \right]} \right\}dx.
\end{eqnarray}
Inserting the generating function of $H_n (x) = \left. {\frac{{\partial ^n }}{{\partial s^n }}e^{2xs - s^2 } } \right|_{s = 0}$ for Hermite polynomial into the above formula, we have
\begin{eqnarray}
W_n \left( {Y,p_y } \right) &=& \frac{{N_n^2 }}{{2\pi }}e^{ - \alpha ^2 Y^2 } \int\limits_{ - \infty }^{ + \infty } \exp \left( { - \frac{\hbar }{4}x^2 } \right)\left. {\left[ {\frac{{\partial ^n }}{{\partial s^n }}\text{e}^{2\alpha (Y - \frac{\hbar }{2}x)s - s^2 } } \right]} \right|_{s = 0} \nonumber\\
&&\left. {\left[ {\frac{{\partial ^n }}{{\partial t^n }}\text{e}^{2\alpha (Y + \frac{\hbar }{2}x)t - t^2 } } \right]} \right|_{t = 0} \text{e}^{ip_y x} dx.
\end{eqnarray}
Namely,
\begin{eqnarray}
W_n \left( {Y,p_y } \right) &=& \frac{{N_n^2 }}{{2\pi }}\text{e}^{ - \alpha ^2 Y^2 } \frac{{\partial ^{2n} }}
{{\partial s^n \partial t^n }}
 \bigg[ \int\limits_{ - \infty }^{ + \infty }
\text{exp} \left( { - \frac{{\alpha ^2 \hbar ^2 }}{4}x} \right)\text{exp} \left( {2\alpha (Y - \frac{\hbar }{2}x)s - s^2 } \right) \nonumber\\
&&\text{exp} \left( {2\alpha (Y + \frac{\hbar }{2}x)t - t^2 } \right)\text{e}^{ip_y x} dx  \bigg] \bigg|_{\scriptstyle s = 0 \hfill \atop
\scriptstyle t = 0 \hfill}.
\end{eqnarray}
That is,
\begin{eqnarray}
W_n \left( {y,p_y } \right) &=& \frac{{N_n^2 }}{{2\pi }}\text{e}^{ - \alpha ^2 Y^2 } \frac{{\partial ^{2n} }}{{\partial s^n \partial t^n }}\left\{ {\exp \left[ {2\alpha Y\left( {s + t} \right) - 2st - \frac{{2ip_y \left( {s - t} \right)}}{{\alpha \hbar }}} \right]} \right. \nonumber\\
&& \left. {\left. {\int\limits_{ - \infty }^{ + \infty } {\exp \left[ { - \frac{{\alpha ^2 \hbar ^2 }}{4}x^2  + \alpha \hbar x\left( {t - s} \right) - s^2  - t^2  + 2st + ip_y x + \frac{{2ip_y \left( {s - t} \right)}}{{\alpha \hbar }}} \right]} } \right\}} \right|_{\scriptstyle s = 0 \hfill \atop
\scriptstyle t = 0 \hfill} \nonumber\\
&=&\frac{{N_n^2 }}{{2\pi }}\text{e}^{ - \alpha ^2 Y^2 } \frac{{\partial ^{2n} }}{{\partial s^n \partial t^n }}
 \bigg\{[\exp \left[ {2\alpha Y\left( {s + t} \right) - 2st - \frac{{2ip_y \left( {s - t} \right)}}{{\alpha \hbar }}} \right] \nonumber\\
&& \int\limits_{ - \infty }^{ + \infty } {\exp \left[ { - \frac{1}{4}\left( {\alpha \hbar x + 2s - 2t} \right)^2  + \frac{{ip_y }}{{\alpha \hbar }}\left( {\alpha \hbar x + 2s - 2t} \right)} \right]dx} \bigg\}  \bigg|_{\scriptstyle s = 0 \hfill \atop
  \scriptstyle t = 0 \hfill}.
\end{eqnarray}
With the integral formula $\int\limits_{ - \infty }^{ + \infty } {\exp \left( { - \beta ^2 x^2  \pm 2i\gamma x} \right)} dx = \frac{{\sqrt \pi  }}{\beta }e^{ - \gamma ^2 /\beta ^2 }$, the above formula can be shifted to
\begin{eqnarray}
W_n \left( {y,p_y } \right) &=& \frac{{N_n^2 \sqrt \pi  }}{{\pi \alpha \hbar }}\exp \left[ { - \left( {\alpha ^2 Y^2  + \frac{{p_y^2 }}{{\alpha ^2 \hbar ^2 }}} \right)} \right] \nonumber\\
&&\left. {\left\{ {\frac{{\partial ^{2n} }}{{\partial s^n \partial t^n }}\exp \left[ {2\alpha y(s + t) - 2st - \frac{{2ip_y }}{{\alpha \hbar }}(s - t)} \right]} \right\}} \right|_{\scriptstyle s = 0 \hfill \atop
  \scriptstyle t = 0 \hfill}.
\end{eqnarray}
With $M_n \equiv \frac{{\partial ^{2n} }}{{\partial s^n \partial t^n }}\exp \left[ {2\alpha Y(s + t) - 2st - \frac{{2ip_y }}{{\alpha \hbar }}(s - t)} \right]$, we have
\begin{eqnarray}
W_n \left( {y,p_y } \right) = \frac{{N_n^2 \sqrt \pi  }}{{\pi \alpha \hbar }}{\mathop{\rm e}\nolimits} ^{ - \left( {\alpha ^2 Y^2  + \frac{{p_y^2 }}{{\alpha ^2 \hbar ^2 }}} \right)}  {M_n } \bigg|_{\scriptstyle s = 0 \hfill \atop
  \scriptstyle t = 0 \hfill}.~\label{eq26}
\end{eqnarray}

By using the induction, we obtain
\begin{eqnarray}
M_n  &=& \sum\limits_{k = 0}^n \frac{{( - 1)^{n + k} (n!)^3 2^{n - k} }}{{(k!)^2 (n - k)!}}\left(2\alpha Y - 2t - \frac{{2ip_y }}{{\alpha \hbar }}\right)^k \left(2\alpha Y - 2s + \frac{{2ip_y }}{{\alpha \hbar }}\right)^k \nonumber\\
&&\text{e}^{2Y(s + t) - 2st - 2ip_y (s - t)/\hbar } .
\end{eqnarray}
Plugging Eq.~(\ref{eq26}) into the above equation, and considering $N_n^2  = \frac{\alpha }{{\sqrt \pi  2^n n!}}$, $W_n \left( {y,p_y } \right)$ can be
written as
\begin{equation}
W_n \left( {Y,p_y } \right) = \frac{{( - 1)^n }}{{\pi \hbar }}\text{e}^{ - (\alpha ^2 Y^2  + \frac{{p_y^2 }}{{\alpha ^2 \hbar ^2 }})/\hbar } \sum\limits_{k = 0}^n {\frac{{( - 1)^k (n!)^2 }}{{(k!)^2 (n - k)!}}\left( {2\alpha ^2 Y^2  + \frac{{2p_y^2 }}{{\alpha ^2 \hbar ^2 }}} \right)^k }.
\end{equation}

Defining $\xi  = 2\alpha ^2 Y^2  + \frac{{2p_y^2 }}{{\alpha ^2 \hbar ^2 }}$, and considering $L_n \left( \xi  \right) = \sum\limits_{k = 0}^n {\frac{{( - 1)^k (n!)^2 }}{{(k!)^2 (n - k)!}}\left( \xi  \right)^k }$ is Laguerre polynomials, we have
\begin{equation}
W_n \left( {y,p_y } \right) = \frac{{( - 1)^n }}{{\pi \hbar }}\text{e}^{ - \xi /2} L_n \left( \xi  \right),
\end{equation}
which is the cross Wigner function for single electron in two-dimensional electron gas system with Hall electric field.
Now we deduce the Wigner function of the longitudinal direction (X), namely,

\begin{eqnarray}
W\left( {x,p_x } \right) &=& \frac{1}{{2\pi L_x }}\int\limits_{ - \frac{{L_x }}{2}}^{\frac{{L_x }}{2}} {\exp \left( {ip_x y} \right)\exp \left[ { - \frac{{ip'_x }}{\hbar }\left( {x + \frac{\hbar }{2}y} \right)} \right]} \exp \left[ {\frac{{ip'_x }}{\hbar }\left( {x - \frac{\hbar }{2}y} \right)} \right]dy \nonumber\\
&=& \frac{1}{{2\pi }}\int\limits_{ - \infty }^{ + \infty } {\exp \left( {ip_x y} \right)\exp \left[ { - \frac{{ip'_x }}{\hbar }\left( {x + \frac{\hbar }{2}y} \right)} \right]} \exp \left[ {\frac{{ip'_x }}{\hbar }\left( {x - \frac{\hbar }{2}y} \right)} \right]dy \nonumber\\
&=& \frac{1}{{2\pi }}\int\limits_{ - \infty }^{ + \infty } {\exp \left( {ip_x y - ip'_x y} \right)} dy \nonumber\\
&=& \delta (p_x  - p'_x ),
\end{eqnarray}
where the second equality is held for $L_x\to \infty$ \cite{18}. Therefore, the Wigner function for integer quantum Hall effect is
\begin{equation}
W = W_{x,p_x } W_{y,p_y }  = \frac{{( - 1)^n }}{{\pi \hbar }}\text{e}^{ - \xi /2} L_n \left( \xi  \right)\delta (p_x  - p'_x ), ~\label{31}
\end{equation}
which represents a significant theoretical achievement with profound physical implications. This formulation provides a complete phase-space characterization of the quantum state distribution for a two-dimensional electron gas under strong magnetic fields.  Importantly, the Wigner function in the above equation is not merely an abstract construct but has direct experimental relevance.  As demonstrated in Ref.~\cite{19}, the Wigner quasi-probability distribution can indeed be experimentally reconstructed by measuring the interference pattern generated in a double-slit experiment with helium atoms. This landmark study confirmed that the Wigner function, despite its mathematical complexity, is a physically measurable quantity that provides a complete description of a quantum system's phase-space properties. The fact that our analytically derived Wigner function for the IQHE (Eq.~(\ref{31})) is expressed in a closed form combining Laguerre polynomials, exponential decay, and delta-function constraints makes it particularly amenable to experimental verification. Specifically, the oscillatory structure introduced by the $(-1)^n L_n(\xi)$ term and the sharp momentum conservation in the $x$-direction ($\delta(p_x-p_x')$) could, in principle, be probed using advanced quantum measurement schemes, such as those employed in Ref.~\cite{19}. Such experimental validation would not only confirm the theoretical predictions but also provide deeper insights into the quantum coherence and topological protection mechanisms underlying the IQHE.

\section{Conclusion}~\label{sec6}

This paper begins by introducing the Wigner function and Landau energy levels, derives the wave function and the Wigner function for the integer quantum Hall effect via the integral method, and finally employs the Wigner function to compute the current density expectation and the quantized Hall resistance formula. While this single-particle, zero-temperature framework successfully captures the essential quantization, its extension to finite temperatures and dissipative regimes is desirable for direct experimental relevance; furthermore, incorporating many-body interactions into the phase-space picture would bridge the gap to fractional Hall physics, and a systematic study of gauge and curvature effects could deepen our understanding of the geometric origin of topological invariants-a connection that remains only implicit in our current derivation and warrants further exploration through semiclassical or topological field-theoretic approaches.

\acknowledgments{
This paper is supported by the National Natural Science Foundation of China (Nos. 11175053 and 11475051), Shaanxi Provincial Department of Education Scientific Research Project (No. 12JK0960), and Natural Science Basic Research Program of Shaanxi (No. 2025JC-YBMS-098). The authors thank Professor Jianhua Wang for his valuable discussions during the preparation of this manuscript. His insights are gratefully acknowledged.}

  %\quad\quad  E-MAILjianhuawang59@aliyun.com   \quad\quad\quad\quad   Tel:13991625329
\end{document}